\documentclass[aps,prd,preprint,superscriptaddress]{revtex4}
\usepackage{longtable}
\usepackage{graphicx}
\usepackage{amsmath,amssymb}
\usepackage{color}
\usepackage{bbm}

\newcommand{\beq}{\begin{equation}}
\newcommand{\eeq}{\end{equation}}
\newcommand{\bdm}{\begin{displaymath}}
\newcommand{\edm}{\end{displaymath}}

\begin{document}

\title{Constraining the gravitational wave energy density of the Universe using Earth's ring}


\author{Michael Coughlin}
\affiliation{Department of Physics, Harvard University, Cambridge, MA 02138, USA}
\author{Jan Harms}
\affiliation{INFN, Sezione di Firenze, Sesto Fiorentino, 50019, Italy}


\date{\today}

\begin{abstract}
The search for gravitational waves is one of today's major scientific endeavors. A gravitational wave can interact with matter by exciting vibrations of elastic bodies. Earth itself is a large elastic body whose so-called normal-mode oscillations ring up when a gravitational wave passes. Therefore, precise measurement of vibration amplitudes can be used to search for the elusive gravitational-wave signals. Earth's free oscillations that can be observed after high-magnitude earthquakes have been studied extensively with gravimeters and low-frequency seismometers over many decades leading to invaluable insight into Earth's structure. Making use of our detailed understanding of Earth's normal modes, numerical models are employed for the first time to accurately calculate Earth's gravitational-wave response, and thereby turn a network of sensors that so far has served to improve our understanding of Earth, into an astrophysical observatory exploring our Universe. In this article, we constrain the energy density of gravitational waves to values in the range 0.035 -- 0.15 normalized by the critical energy density of the Universe at frequencies between 0.3\,mHz and 5\,mHz, using 10 years of data from the gravimeter network of the Global Geodynamics Project that continuously monitors Earth's oscillations. This work is the first step towards a systematic investigation of the sensitivity of gravimeter networks to gravitational waves. Further advance in gravimeter technology could improve sensitivity of these networks and possibly lead to gravitational-wave detection.
\end{abstract}

\pacs{04.80.Nn,91.30.Fn,95.75.Wx}

\maketitle 

\section{Introduction}
So far, the strongest evidence for the existence of gravitational waves (GWs) comes from the observation of the binary pulsar PSR B1913+16 \cite{WNT2010}. The shrinking of its orbit observed over three decades can be fully explained by the emission of GWs and associated energy loss according to the General Theory of Relativity. Dedicated experiments attempt to measure these waves as phase modulation of laser beams (GEO600 \cite{LuEA2010}, LIGO \cite{LSC2010}, Virgo \cite{Vir2011}, KAGRA \cite{AsEA2013}, eLISA \cite{eLISA2012}, TOBA \cite{AnEA2010b}), or through their imprint on the polarization of the cosmic microwave background (BICEP2 \cite{BrEA2010}, EBEX \cite{OxEA2004}). Furthermore, GWs can be searched in data of other high-precision experiments including Doppler tracking of satellites \cite{ArEA2003}, monitoring arrival times of pulsar signals \cite{Hob2005}, or using the Global Positioning System \cite{ATI2014}. Gravitational waves can also excite oscillations of elastic bodies. This principle is exploited for example in the design of spherical resonant GW detectors (MiniGRAIL \cite{WaEA2003}, Mario Schenberg \cite{AgEA2006}). Also oscillations of stars can be excited, and therefore observation of these modes can be used to detect GWs \cite{SiRo2014}. All these experiments combined monitor a wide range of GW frequencies starting from waves that have oscillated only a few times since the beginning of the Universe, up to a few 1000\,Hz.

Recently, the authors of this article have presented results from an observation of the free, flat surface response of Earth to GWs \cite{CoHa2014}. As was explained there, the method cannot be extended to frequencies below about 50\,mHz since seismic motion starts to be globally coherent at lower frequencies, and the GW response is strongly affected by Earth's spherical shape. The low-frequency GW response is best described in terms of Earth's normal-mode oscillations \cite{Ben1983}. These oscillations are continuously monitored by a global network of low-frequency seismometers and gravimeters. Especially the superconducting gravimeters of the Global Geodynamics Project (GGP), which were used in this paper, provide excellent sensitivity below 10\,mHz with data records reaching back more than 10 years \cite{CrHi2010}. As will be shown, the stationary noise background is almost the same for all gravimeters and uncorrelated between different instruments, which makes it possible to use a large fraction of the data of the entire network to search for GW signals that are significantly weaker than the stationary noise level by means of a near-optimal correlation method. Whereas previous GW searches using Earth's normal modes only tried to explain excess energy in normal modes \cite{WeBl1965,Tum1971}, the work in this article is the first to combine a near-optimal analysis of gravimeter data with a detailed GW response model, which makes it possible to accurately calibrate normal-mode amplitudes into GW strain. 

\begin{figure}
\includegraphics[width=0.4\textwidth]{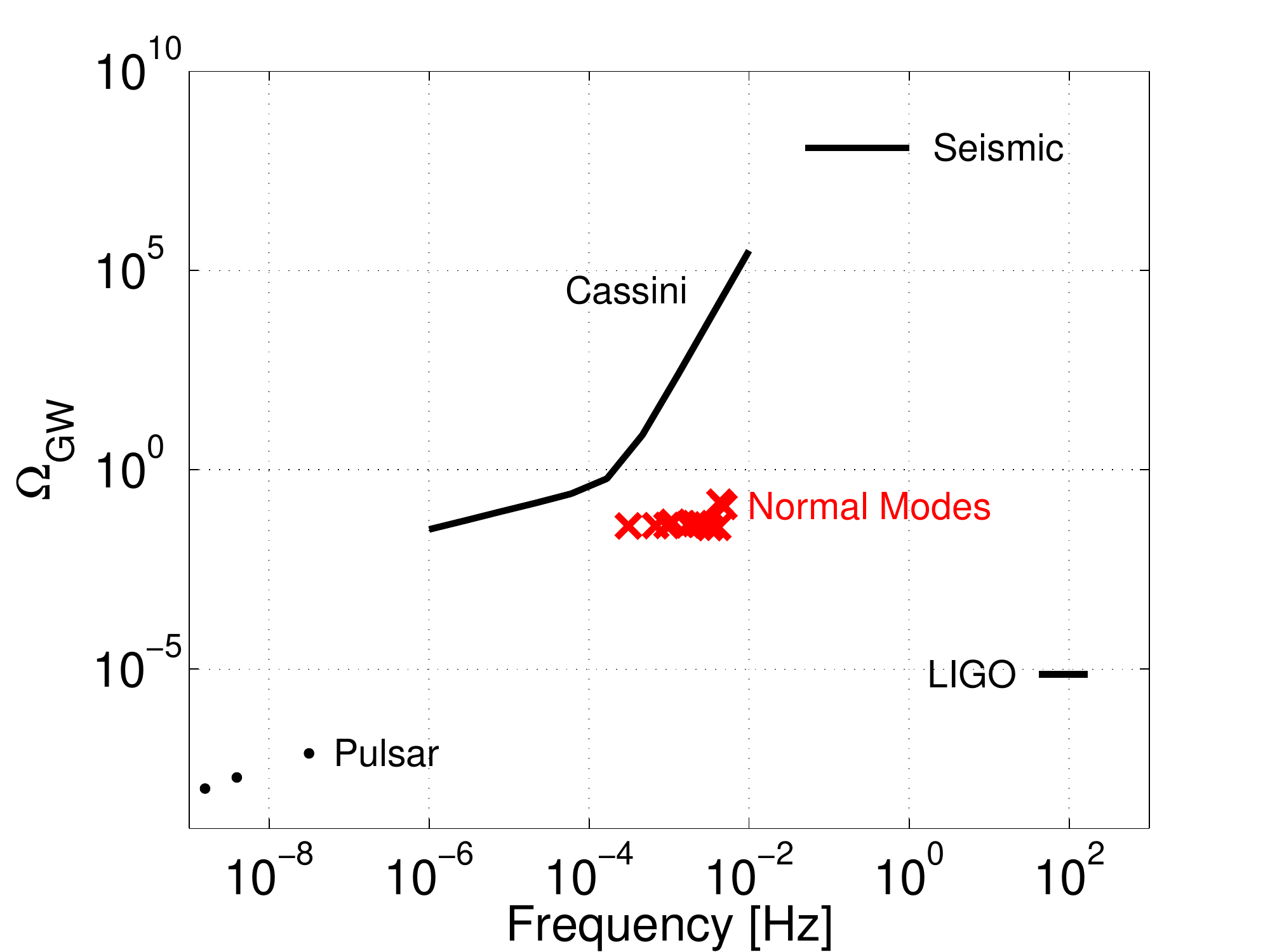}
\caption{Current upper limits on GW energy density. These limits were set by pulsar timing observations \cite{JeEA2006}, Doppler-tracking measurements of the Cassini spacecraft \cite{ArEA2003}, monitoring Earth's free-surface response with seismometers (``Seismic'') \cite{CoHa2014}, and correlating data from the first-generation, large-scale GW detectors LIGO \cite{AbEA2009}. The new limits resulting from normal-mode measurements are shown as crosses.}
\label{fig:upperlimits}
\end{figure}

The limits obtained in this study through normal-mode observations are plotted in Fig.~\ref{fig:upperlimits} together with upper limits set in other frequency bands. The previous upper limits in the frequency range 0.3\,mHz to 5\,mHz are improved by 2 to 5 orders of magnitude. 
A brief summary of normal-mode oscillations is given in section \ref{sec:normal}. In section \ref{sec:theory}, we outline the theory of Earth's resonant (normal-mode) response to GWs. A characterization of gravimeter data is presented in section \ref{sec:gravimeter}. Finally, the GW search algorithm is discussed in section \ref{sec:search}, and new constraints are presented on the energy density of GWs averaged over directions and wave polarizations.

\section{Earth's normal-mode oscillations}
\label{sec:normal}
Earth's free oscillations, called normal modes, can be excited by gravitational waves. Earth's slowest normal-mode oscillation occurs at about 0.3\,mHz, and distinct modes can still be identified up to a few millihertz. At higher frequencies, the discrete vibrational spectrum transforms into a quasi-continuous spectrum of seismic vibrations that are increasingly dominated by local sources. The data used in this study were sampled once per minute, and low-pass filtered suppressing signal response above about 5\,mHz depending on the gravimeter. In addition, a few gravimeters show resonant features above 5\,mHz in their response. Therefore, the upper frequency bound of the GW search was chosen to be 5\,mHz to guarantee accurate calibration of the data.
 
At frequencies below 5\,mHz, the diameter of Earth is much smaller than the length of GWs. In this so-called long-wavelength regime, a GW can effectively be represented by a quadrupole-force field that excites Earth's normal modes. Normal modes are divided into toroidal $_nT_l$ and spheroidal $_nS_l$ modes, where $n,l$ are non-negative integers that determine the radial and angular mode shape respectively. The toroidal modes only produce tangential displacement. Spheroidal modes show tangential and radial displacement, and they also perturb Earth's gravity field. Not all normal modes are equally responsive to a quadrupole force. In fact, only the quadrupole modes with $l=2$ show significant GW response in the long-wavelength regime \cite{Ben1983}. The coupling mechanism of a GW to oscillations of elastic bodies is governed by variations of the shear modulus, including the shear-modulus change across the free surface. Earth shows strong internal variations of the shear modulus. In the liquid outer core, the shear modulus vanishes, and therefore significant internal contributions to Earth's GW response can be expected at the inner-core boundary, as well as at the core-mantle boundary. Due to the complex internal structure of Earth, normal modes also show a complex radial dependence of their amplitudes. Modes with the high amplitudes at the inner-core boundary, core-mantle boundary, and free surface couple strongly to GWs.

In order to calculate the response of Earth to GWs, normal-mode amplitudes as a function of radius need to be modelled numerically. For superconducting gravimeters, three contributions need to be modelled and added coherently: seismic acceleration, perturbation of the gravity potential, and lift against a static gravity gradient. For this work, normal-mode solutions were generated with the numerical simulation tool Minos \cite{Woo1988}. These solutions are valid for a spherical, non-rotating, laterally homogeneous Earth, and here are based on the Earth model PREM \cite{DzAn1981} that describes variations of mass density, seismic speeds, and damping parameters from Earth's center to its surface. The gravimeters are designed to measure radial ground motion and gravity changes, which are caused only by spheroidal modes. Therefore, one can focus on these modes for the GW search. Of all spheroidal quadrupole modes $_nS_2$, only 14 have frequencies $f_n$ below 5\,mHz as shown in Table \ref{tab:params}. Even though Earth also responds to GWs off these 14 mode frequencies, the best sensitivity is obtained at normal-mode frequencies making use of the resonant signal amplification. The GW response at normal-mode frequencies needs to take into account the damping experienced by each mode in order to obtain the correct signal amplification. The damping is quantified by a mode's quality factor, which corresponds to the ratio of a mode frequency to its natural spectral linewidth. The quality factors of the 14 modes lie between about $Q=100$ and 900. The mode frequencies and quality factors used here were all taken from the numerical simulation, but it should be emphasized that numerical estimates of the mode frequencies are very accurate, at least for the purpose of this paper, and also the quality factors agree well with observation \cite{GiDz1975,DaTr1998}.

The coupling strength $\alpha_n$ of a mode to a GW can be expressed by a dimensionless quantity. Its values for the 14 quadrupole modes below 5\,mHz are summarized in Table \ref{tab:params}. They depend on the radial as well as tangential displacement of each mode, and also on shear-modulus changes and mass density as functions of the distance to Earth's center. The coupling strength varies by more than an order of magnitude without clear pattern. This is owed to the complexity of mode solutions, which have greatly varying sensitivity to shear-modulus changes at different depths.
\begin{table*}[t]
\footnotesize
\begin{tabular}{l|p{1cm}|p{1cm}|p{1cm}|p{1cm}|p{1cm}|p{1cm}|p{1cm}|p{1cm}|p{1cm}| p{1cm}|p{1cm}|p{1cm}|p{1cm}|p{1cm}}
$_nS_2$ & 0 & 1 & 2 & 3 & 4 & 5 & 6 & 7 & 8 & 9 & 10 & 11 & 12 & 13 \\
\hline\hline
$f_n$ [mHz] & 0.309 & 0.679 & 0.938 & 1.11 & 1.72 & 2.09 & 2.41 & 2.52 & 3.21 & 3.23 & 4.03 & 4.06 & 4.33 & 4.84 \\
$Q_n$ & 510 & 310 & 95.9 & 365 & 433 & 317 & 92.9 & 340 & 316 & 445 & 203 & 126 & 229 & 878 \\
$\alpha_n$ & $-$0.645 & $-$18.3 & $-$1.78 & $-$0.696 & $-$18.9 & 13.3 & 4.31 & $-$34.5 & $-$3.97 & $-$6.54 & 15.8 & $-$16.9 & 12.7 & 3.12 \\
$u_n$ &  0.74 & -0.14 & -0.06 & -0.19 & -0.18 & -0.11 & -0.019 & -0.086 & -0.050 & 0.13 & 0.073 & 0.057 & -0.021 & 0.086 \\
$p_n$ & -0.43 &  0.028 & 5.7e-5 & -4.9e-4 & 2.1e-4 & 4.7e-5 & -3.1e-6 & 1.7e-5 & 5.3e-6 & 3.6e-6 & 2.0e-7 & 7.4e-7 & 1.0e-6 & 1.9e-8 \\
$\Omega_{\rm GW}$ & 0.039 & 0.039 & 0.040 & 0.048 & 0.041 & 0.045 & 0.042 & 0.044 & 0.035 & & 0.036 & & 0.15 & 0.12 \\
\hline
\end{tabular}
\caption{Summary of mode parameters: mode frequencies $f_n$, quality factors $Q_n$, coupling strengths $\alpha_n$, radial surface displacement $u_n$, perturbation of gravity surface potential $p_n$ (both normalized to the same, but arbitrary unit). The last row shows the upper limits on the energy density $\Omega_{\rm GW}$ as plotted in Fig.~\ref{fig:pointestimates}.}
\label{tab:params}
\end{table*}
In addition to the coupling strengths, the second important parameter characterizing each mode is its vertical displacement $u_n$ and gravity potential perturbation $p_n$ at the surface, which govern the gravimeter signal. These amplitudes are also summarized in Table \ref{tab:params}. The amplitudes of displacement and gravity potential are normalized such that their relative contribution to the gravimeter signal can be compared. It can be seen that the gravity perturbation is significant only for the two modes $_0S_2$ and $_1S_2$.

A feature of normal modes that is not captured by the Minos simulation is mode coupling due to Earth's ellipticity, rotation, and lateral heterogeneity. One effect is the so-called self coupling, in which a quadrupole ($l=2$) multiplet can split into up to 5 resolvable modes, which are labelled by a third integer $m=-2,\ldots,2$ \cite{HeTr1996}. Since each mode can therefore potentially respond to a different GW, mode splitting influences the overall GW response. Another possibility is that two modes that happen to be very close in frequency can couple and exchange energy. 
\begin{figure}
\includegraphics[width=0.4\textwidth]{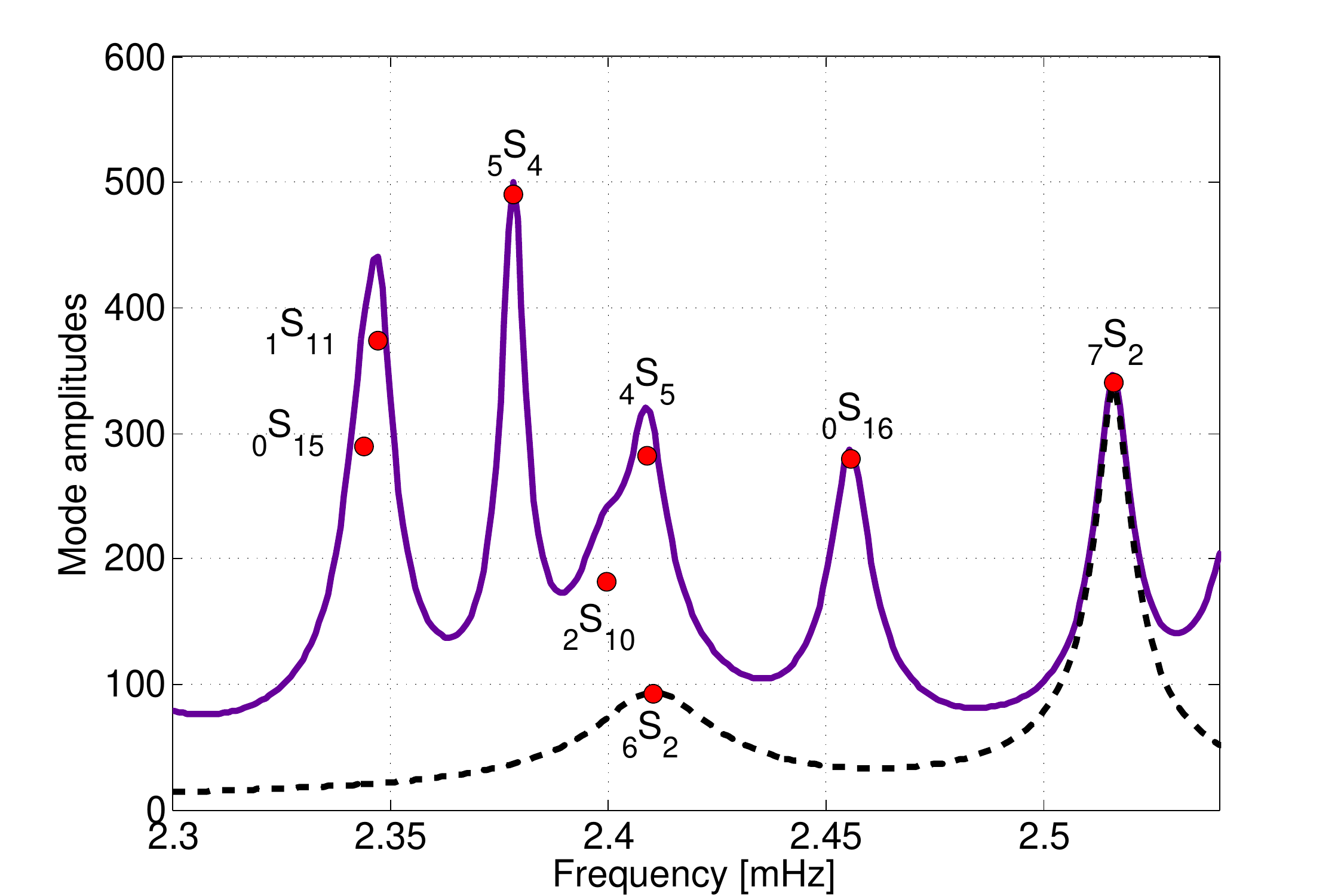}
\caption{Simulated spectrum of spheroidal normal modes around $_6S_2$. The values of the red markers correspond to the modes' $Q$-factors.}
\label{fig:multiplet}
\end{figure}
The latter situation is depicted in figure \ref{fig:multiplet} taking the mode $_6S_2$ as example. Whereas the next highest quadrupole mode $_7S_2$ is well isolated, mode $_6S_2$ lies very close to other spheroidal modes, which can couple and exchange energy. Although the effect is minor on normal-mode frequencies and $Q$-values \cite{Dah1968,IDW2009}, a consequence is that one cannot design the GW search into too narrow frequency bands only relying on simulation predictions. An extreme narrow-band search needs to be based on a detailed characterization of the quadrupole modes taking into account observed mode (self-)coupling, which has not been done in the work presented here. Concerning energy transfer between coupled modes, the effect would generally lead to a decrease in GW response of a quadrupole mode independent of the $Q$-values of the coupled modes. However, estimating the change in GW response that is consistent with observed shifts of normal-mode frequencies (based on a simple coupled harmonic oscillator model), it can be concluded that the energy lost into other modes through coupling is negligible. Therefore, the main issue with mode coupling is that the GW search needs to be designed with sufficient bandwidth around each mode frequency so that it is guaranteed that the peak response of the entire quadrupole multiplet lies within this band. Further details about the impact of mode coupling on GW sensitivity are given in section \ref{sec:search}.

\section{Theory of Earth's response to gravitational waves}
\label{sec:theory}
Two response mechanisms of an elastic body to GWs have been described in detail in past publications. First, Dyson calculated the amplitude of seismic waves produced by GWs incident on a free, flat surface \cite{Dys1969}. He found that the first time derivative of vertical surface displacement is given by
\beq
\dot\xi_z(\vec r\,,t)\approx-\frac{\beta^2}{\alpha}\vec e_z^\top\cdot h(\vec r\,,t)\cdot \vec e_z
\label{eq:flatresp}
\eeq
Here, $\vec e_z$ denotes the normal vector of the surface, $h$ the spatial part of the GW strain tensor, and $\alpha,\,\beta$ are the compressional and shear-wave speed. It can already be seen that the shear modulus $\mu$ plays an important role in the elastic-body response since
\beq
\beta^2=\frac{\mu}{\rho}.
\eeq 
Accordingly, the GW response vanishes for vanishing shear modulus. One has to keep in mind though that the equations of elastic deformation used to derive this result are neglecting contributions that can become important when the shear modulus is sufficiently small. For example, the GW response model of a spherical body with vanishing shear modulus has been used by Siegel and Roth \cite{SiRo2014} to propose GW measurements by monitoring oscillations of the Sun.

The role of the shear modulus is also evident in the GW response of an elastic spherical body. This case was studied by Ben-Menahem \cite{Ben1983}, and is used here to calculate Earth's resonant GW response. In the following, we will present the most important results of his work with minor reformulations. In terms of the amplitudes of radial displacement $u_{n2m}(r)$, and tangential displacement $v_{n2m}(r)$, the coupling strength of a GW to a normal quadrupole mode can be defined as
\begin{equation}
\alpha_{n2m}\equiv -\frac{R}{\beta_{\rm c}^2}\frac{\int_0^{R+}{\rm d}r\,r^2\mu'(r)(u_{n2m}(r)+3v_{n2m}(r))}{\int_0^R{\rm d}r\,r^2\rho(r)(u_{n2m}^2(r)+6v_{n2m}^2(r))},
\label{eq:sphereresp}
\end{equation}
where $R$ is Earth's radius, $\beta_{\rm c}$ is the shear-wave speed at Earth's center, $\mu'(r)$ the derivative of the shear modulus, and $\rho(r)$ the mass density. The upper integration limit $R+$ signifies that the shear-modulus change across the free surface needs to be included. In the following, only the radial order $n$ will be used to specify a quadrupole mode whose properties are independent of the index $m$ neglecting mode coupling. The mode amplitudes $u_n(r),\,v_n(r)$ have arbitrary units, since units of the mode variables cancel in the final result. They are considered unitless in this work. It is only necessary that all mode variables including the amplitude $\phi_n(r)$ of the gravity potential are normalized consistently.

The complete solution for the GW response also needs to take into account the angular dependence of excited oscillations. A simple first step is to consider the response to a single, plus-polarized GW. For a spherical, laterally homogeneous Earth, the acceleration $a_{n2m}$ measured by a gravimeter in the long-wavelength regime can be written
\begin{equation}
\begin{split}
a_{n2m}(&f_n;\theta,\phi) = \frac{\sqrt{24\pi}}{15}\frac{\beta_{\rm c}^2}{R}Q_n\alpha_nh(f_n)\delta_{|m|,2}Y_2^{m\,*}(\theta,\phi)\\
&\cdot\left(u_n(R)+3\frac{\phi_n(R)}{R(2\pi f_n)^2}+2\frac{g}{R(2\pi f_n)^2}u_n(R)\right),
\end{split}
\label{eq:GWresp}
\end{equation}
where $h(f_n)$ is the GW strain amplitude, $g=9.81\,$m/s$^2$, and $\delta_{kl}$ the Kronecker delta. For a quadrupole mode with $l=2$, the angular parameter can take the values $m=-2,\ldots,2$. The expression in the brackets comprises the three contributions to the gravimeter signal: radial surface displacement, perturbation of the gravity potential, and lift against a static gravity gradient \cite{CHR2013}. The second contribution corresponds to the parameter $p_n$ in Table \ref{tab:params}: $p_n\equiv 3\phi_n(R)/(R(2\pi f_n)^2)$. The angle $\theta$ denotes the relative angle between the direction of propagation of the GW and the location of the gravimeter on Earth's surface in a coordinate system with origin at the center of the Earth. The angle $\phi$ describes the rotation of this coordinate system with respect to the polarization frame of the GW. Accordingly, two modes, $m=\pm 2$, of the quadrupole multiplet are excited by each GW in this choice of coordinate system. 

\begin{figure}
\includegraphics[width=0.4\textwidth]{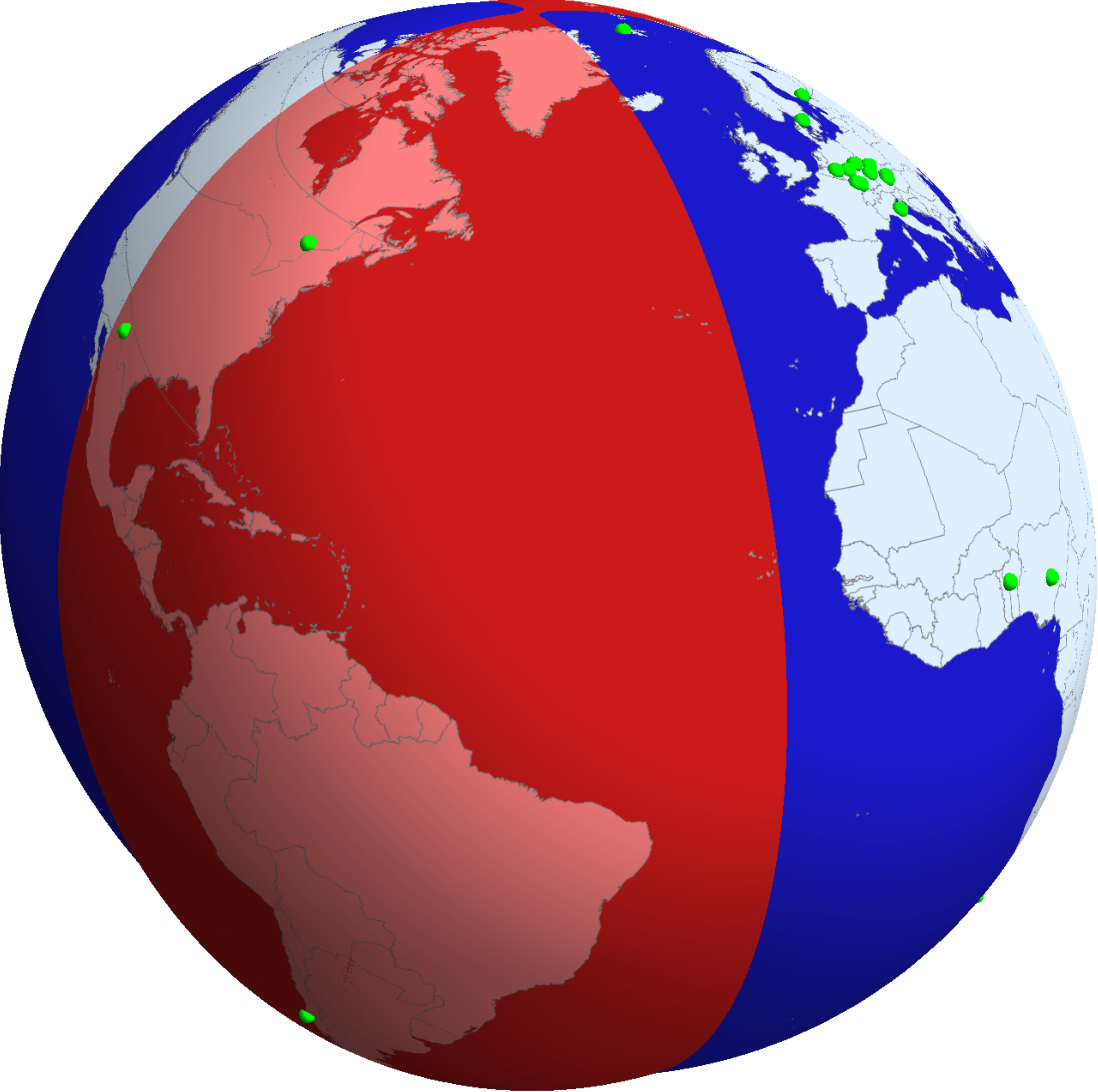}
\caption{Earth quadrupole oscillation. The red and blue shapes correspond to the maxima of a quadrupole oscillation separated by half an oscillation period. The green balls mark locations of some of the gravimeters of the GGP network. Here the oscillation is induced by a GW propagating along the north-south axis.}
\label{fig:quadoscillation}
\end{figure}

For the GW search carried out in this study, we also need to know the correlation between two gravimeters due to an isotropic GW background. Each GW that couples to quadrupole normal modes produces an angular surface vibration pattern that can, in an arbitrarily oriented Earth-centered coordinate system, be represented by a linear combination of quadrupole spherical harmonics $Y_2^m(\theta,\phi)$ with $m=-2,\ldots,2$ \cite{RoKr2007}. The situation is illustrated in Fig.~\ref{fig:quadoscillation}. A plus-polarized GW propagates parallel to the north--south axis. The red and blue colored shapes represent Earth's induced quadrupole oscillation at its two maxima separated by half an oscillation period.
Since the signal amplitude measured by gravimeters depends on their location, coherence between two gravimeters also depends on location. For symmetry reasons, it is clear that for an isotropic GW field, coherence integrated over all polarizations and propagation directions only depends on the relative position of the two gravimeters. This correlation function is known as overlap-reduction function, and normalized such that it is unity for collocated gravimeters \cite{Chr1992}. In order to calculate it, the response as given by equation (\ref{eq:GWresp}) needs to be calculated in a rotated coordinate system for one of the gravimeters. Since the GW correlation also depends on the nature of the GW field, a specific model needs to be chosen. Results in this paper are calculated for an isotropic, stationary field of GWs. Integration over all GW propagation directions and polarizations yields the overlap-reduction function:
\beq
\gamma_{12}(\sigma)=\sqrt{4\pi/5}\, Y_2^0(\sigma,0),
\label{eq:ORF}
\eeq
where $\sigma$ is the angle subtended by the great circle that connects the two gravimeters. All else being equal, the gravimeter pairs that contribute most significantly to the estimate of a GW energy density are either close to each other or antipodal. Note that the overlap-reduction function can be approximated as frequency independent since the Earth is orders of magnitude smaller than the length of a GW at mHz frequencies.

\section{Gravimeter data}
\label{sec:gravimeter}
\begin{figure}
\includegraphics[width=0.45\textwidth]{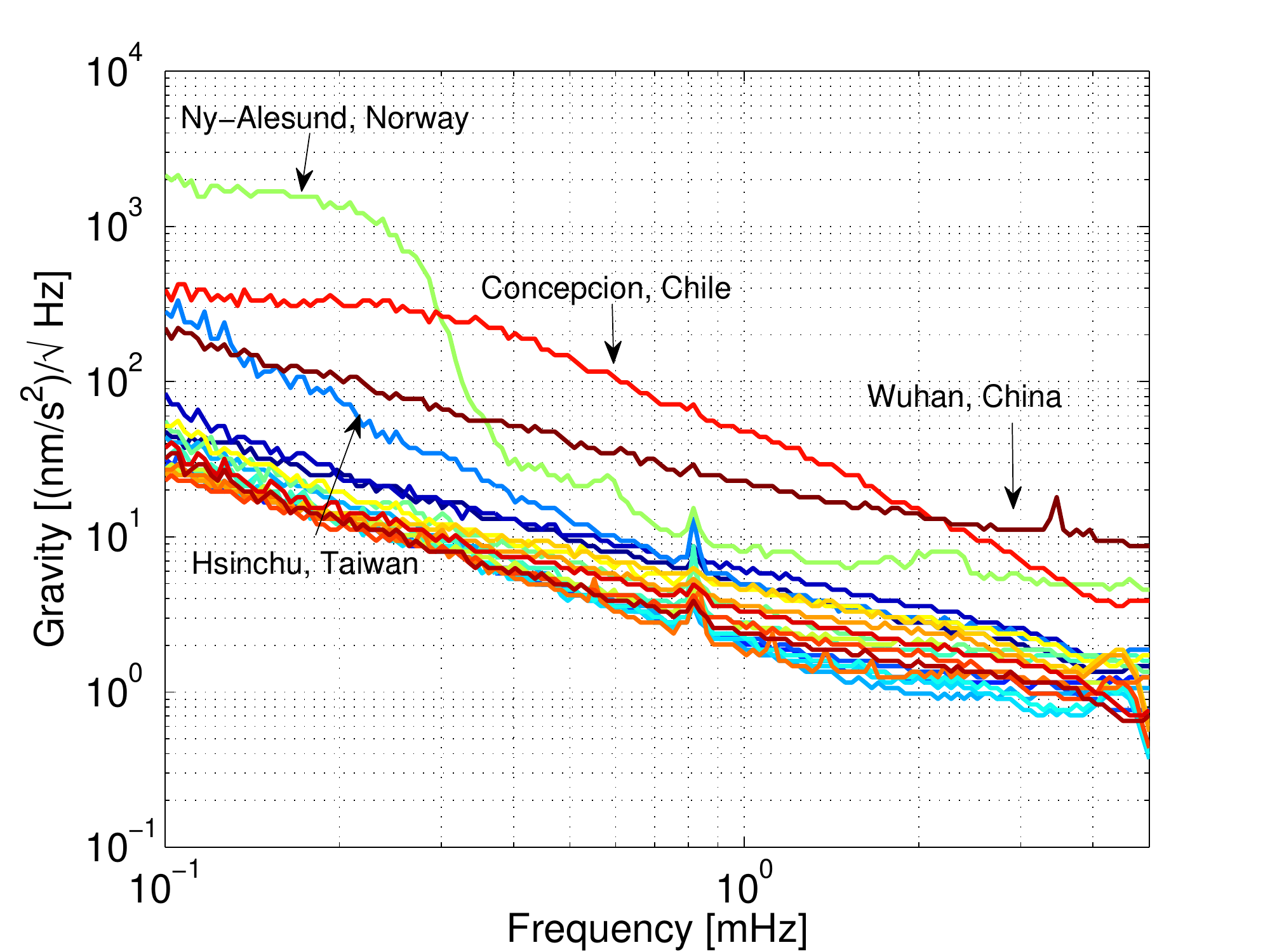}
\caption{Medians of gravimeter spectra measured in 2012. All gravimeters used in this study show a comparable level of stationary background noise represented by their spectral medians, except for the 4 gravimeters highlighted in the plot.}
\label{fig:mediannoise}
\end{figure}

In addition to disturbances from large earthquakes including the subsequent ringdown of the normal modes \cite{XuEA2008}, or local short-duration disturbances, gravimeter data also contain a stationary noise background consisting of instrumental noise, hydrological, and atmospheric disturbances \cite{CHR2013}. The stationary noise level is very similar in almost all instruments, with a median of a few (nm/s$^2$)/Hz$^{1/2}$ at 1\,mHz. The medians of gravimeter spectra recorded during the year 2012 are plotted in Fig.~\ref{fig:mediannoise}.
Four gravimeters show elevated medians, but in all these cases it is not the stationary background being higher, but instead the four instruments are frequently perturbed by strong local events, which therefore contribute significantly to the medians. A detailed study of gravimeter noise for most of these sites can be found in \cite{RoHi2011}.

A local disturbance can produce strong broadband noise in gravimeters. Consequently, noise amplitudes at different frequencies show partial correlation. This property was exploited to subtract some of the background noise that adds to the normal-mode signals, and thereby improve sensitivity to GWs. In this way, it was possible to suppress the background noise at normal modes up to a factor 3 (varying in time, and with different success for each normal mode). Using off-resonance amplitudes for noise subtraction, it is possible to ensure that an insignificant amount of GW signal is subtracted with the noise. Additional noise reduction can be achieved in some gravimeters by direct subtraction of gravity noise of atmospheric origin \cite{Neu2010}. For this purpose, each superconducting gravimeter is equipped with a pressure sensor. The idea is that the pressure data contain direct information about corresponding atmospheric density and therefore gravity perturbations. It is found that the correlation between pressure and gravimeter data is significant below about 1\,mHz and weakly frequency-dependent. This can be exploited to coherently subtract gravity noise with a conversion factor around $-0.35\,\mu$gal/hPa, which needs to be optimized for each gravimeter. The quality of pressure data is poor at some gravimeter sites so that good noise reduction cannot be generally achieved.

\begin{figure}
\includegraphics[width=0.45\textwidth]{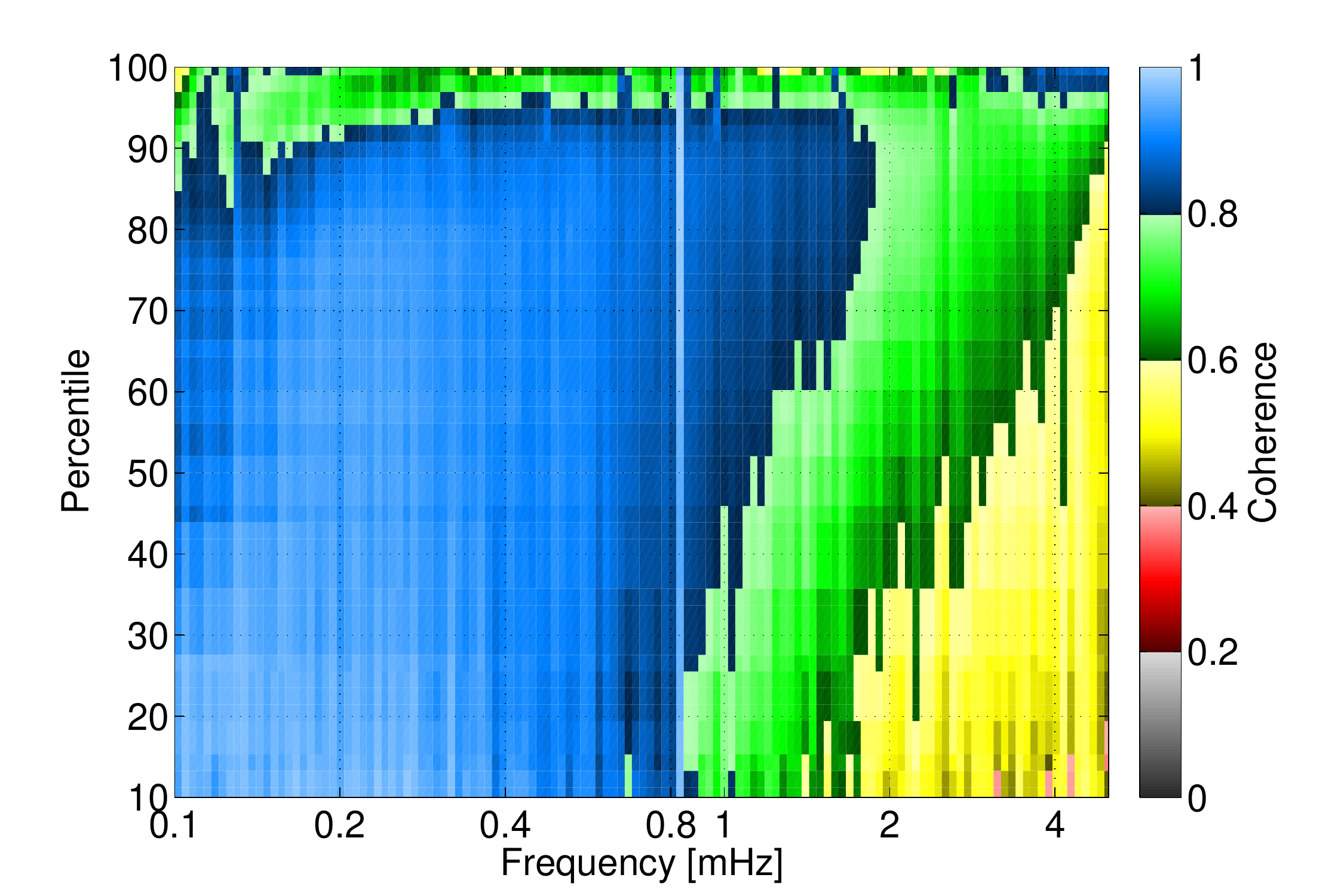}
\caption{Coherence of signals from two levitated spheres in the same gravimeter at Wettzell, Germany. The result is shown as a function of percentile of gravimeter noise excluded from the coherence measurement. A percentile of 90 means that 10\% of the loudest spectra were excluded from the coherence measurement. Only the high-Q radial normal mode $_0S_0$ at about 0.81\,mHz contributes significantly to coherence for all times.}
\label{fig:coherence}
\end{figure}

At should be emphasized that environmental disturbances can show strong correlation well below 0.3\,mHz. Another important property of gravimeter data is that coherence between any two gravimeters of the GGP network at frequencies between 0.3\,mHz and 5\,mHz produced by environmental disturbances is insignificant provided that times of high-magnitude earthquakes are excluded. Even for superconducting gravimeters that contain two levitated spheres, strong coherence is only observed below about 2\,mHz after removing the highest 10th percentile of loud events as shown in Figure \ref{fig:coherence}.
The lack of environmental coherence is an important feature of the gravimeter network, which makes it a very efficient tool to search for GWs, since significant correlations of environmental origin would greatly limit the network sensitivity. 

\section{Search for a stationary gravitational-wave background}
\label{sec:search}
In this section, we outline the GW search method based on correlation measurements between gravimeter pairs. Depending on the relative position of two gravimeters on Earth's surface, correlation of gravimeter signals arising from GWs is described by the overlap-reduction function in equation (\ref{eq:ORF}). Once the expected correlation of GW signals between different gravimeters is calculated, the measured correlations are used to obtain an estimate of the energy density of GWs following the method described in \cite{AlRo1999}. The upper limit on the GW energy density presented in this paper was obtained as a near-optimal combination of measured correlations using 10 years of data, forming pairs with gravimeters of the GGP network. The total amount of data is divided into stretches short enough so that the spectral resolution is wider than the frequency spread of a quadrupole multiplet as discussed in section \ref{sec:normal}. The length of data stretches obtained in this way is different for each normal mode. Each data stretch leads to a point estimate of the GW energy density according to
\begin{equation}
\hat \Omega_{\rm GW}(f_n) = \frac{4\pi^2}{3H_0^2}\frac{\hat S_{12}(f_n)f_n^3}{\gamma_{12}}
\end{equation}
Here, $\hat S_{12}(f_n)$ is the measured cross-spectral density between two gravimeters in units of GW strain spectral density. As pointed out before, the overlap-reduction function $\gamma_{12}$ can be approximated as frequency independent for normal-mode observations.

Based on the conservative assumption that the $l=2$ quadrupole mode splits into 5 distinct isolated modes ($m=-2,\ldots,2$) that all respond incoherently to GWs, the GW response of a quadrupole mode is obtained by adding contributions from different values of $m$ incoherently. Furthermore, two pairs of the 14 quadrupole modes are too close in frequency to be resolvable with the chosen frequency resolution ($n=8,9$ and $n=10,11$, see Table \ref{tab:params}). This means that in addition to the incoherent sum over a multiplet, contributions from the two quadrupole modes in each of these pairs need to be summed incoherently leading to a combined point estimate.

\begin{figure}
\includegraphics[width=0.5\textwidth]{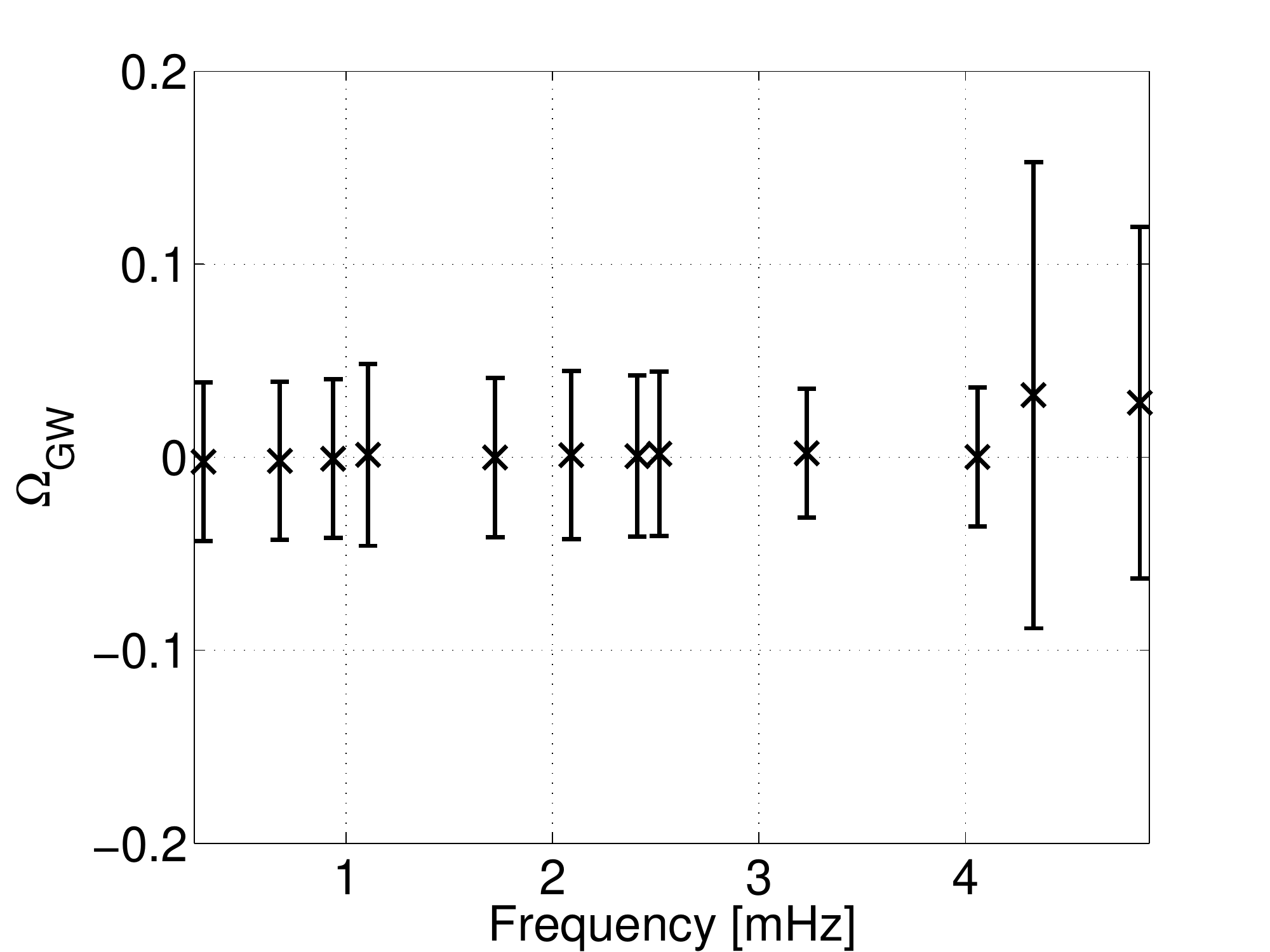}
\caption{Point estimates of GW energy density and errors. Of the 14 original modes, only 12 are plotted here since two pairs, $n=8,9$, and $n=10,11$, have been merged to one value each since the chosen frequency resolution cannot resolve them.}
\label{fig:pointestimates}
\end{figure}

The final results will be presented as constraints on the energy density in GWs separately for each mode. Figure \ref{fig:pointestimates} shows the point estimates of the GW energy density with error bars. All point estimates are consistent with a non-detection, and the resulting energy constraints are mostly determined by the error bars. The values are listed in Table \ref{tab:params}.
Energy densities can be translated into strain spectral densities, which lie between $h_{\rm GW}\leq 2.2\times 10^{-14}\, \rm Hz^{-1/2}$ for the mode $_0S_2$ and $h_{\rm GW}\leq 6.2\times 10^{-16}\, \rm Hz^{-1/2}$ for $_0S_{13}$. Even though these results demonstrate an improvement in sensitivity by a few orders of magnitude over previous searches in this frequency band (see Fig.~\ref{fig:upperlimits}), the new upper limits are still not stringent enough to constrain cosmological models of GW backgrounds. A conservative estimate of the energy density of GWs from inflation predicts a value of order $\Omega_{\rm GW}\sim 10^{-15}$, and a GW background from cosmic strings is predicted at $\Omega_{\rm GW}\sim 10^{-7}$, both at normal-mode frequencies \cite{AbEA2009}. Also a GW background from a cosmological distribution of unresolved compact binary stars such as white dwarfs and neutron stars is predicted at lower values, $\Omega_{\rm GW}\sim 10^{-12}$, at normal-mode frequencies \cite{FaPh2003}. Therefore, with achieved upper limits between $\Omega_{\rm GW}=0.035$ -- $0.15$, the stationary gravimeter noise as plotted in Fig.~\ref{fig:mediannoise} has to be lowered by 2 to 3 orders of magnitude to be able to place first constraints on cosmological models. 

\section{Conclusion}
In this paper we showed that today's understanding of Earth's interior can be used to accurately calculate Earth's resonant GW response. In this way, it was possible to calibrate gravimeter data into units of GW strain, and directly obtain new upper limits on the GW energy density in the range 0.035 -- 0.15 at frequencies between 0.3\,mHz and 5\,mHz. This was achieved by correlating data of gravimeter pairs recorded over the past 10 years. 

Alternatively, one could make use of the same response mechanism to search for individual astrophysical signals such as galactic white-dwarf binaries. Millions of binaries are predicted to radiate quasi-monochromatic waves in this frequency band \cite{CrCo2007} including already discovered systems (see for example Roelofs et al \cite{RoEA2010}). Again, about 3 orders of magnitude sensitivity improvement are required to make a detection likely. The integrated gravitational-wave signal should be distinguishable from terrestrial sources since it is modulated due to Earth's rotation. The additional challenge here is that a continuous integration of the signal results in an extremely narrow frequency resolution, which requires a more detailed investigation of mode-coupling effects. The diversity in nature of Earth's oscillations also makes it possible to test alternative theories of gravity. For example, a scalar component of the GW field could be searched in monopole modes $_nS_0$ as has already been attempted by Weiss and Block \cite{WeBl1965}. 

Further improvement in GW sensitivity may be achieved with a new generation of gravimeters. Especially atom-interferometric gravimeters are currently under active development \cite{DiEA2013}. The open question is if there will be some form of environmental noise limiting the sensitivity of gravimeters irrespective of their intrinsic acceleration sensitivity, and whether methods can be developed to mitigate this noise if necessary. Nonetheless, we have demonstrated that gravimeter technology is a viable option to detect GWs, and that ground-based GW detection seems to be a possibility at frequencies, which are generally considered accessible only for space-borne detectors.

\section{Acknowledgments}
We want to thank David Crossley for helpful discussions on gravimeter data, and Jean-Paul Montagner, Guy Masters, and Eric Cl\'ev\'ed\'e for their help on normal-mode simulations. MC was supported by the National Science Foundation Graduate Research Fellowship
Program, under NSF grant number DGE 1144152. Uncorrected gravimeter data (GGP-SG-MIN) used for this project were downloaded from \url{http://isdc.gfz-potsdam.de/index.php?module=pagesetter&func=viewpub&tid=1&pid=54}.

\raggedright

\end{document}